\documentclass[10pt, a4paper, conference]{IEEEtran}

\ifCLASSINFOpdf
   \usepackage[pdftex]{graphicx}

   \DeclareGraphicsExtensions{.pdf,.jpeg,.png}
\else

\fi

\usepackage{amsmath}

\usepackage{algorithmic}
\usepackage[linesnumbered, ruled]{algorithm2e}

\usepackage{hyperref}

\usepackage{array}

\usepackage{float}
\usepackage{subfig}
\usepackage{xcolor}
\usepackage{threeparttable} 
\usepackage{multirow}
\usepackage{pifont}
\usepackage{diagbox} 
\usepackage{rotating}
\usepackage{booktabs}
\usepackage{tikz}

\newcommand*\halfcirc[1][0.75ex]{%
  \begin{tikzpicture}
  \draw[fill] (0,0)-- (90:#1) arc (90:270:#1) -- cycle ;
  \draw[thick] (0,0) circle (#1);
  \end{tikzpicture}}
\newcommand*\fullcirc[1][0.75ex]{\tikz\fill (0,0) circle (#1);}

\usepackage{tabularx}
\usepackage{mathptmx}  
\usepackage{eucal}
\usepackage{amssymb}

\usepackage{eso-pic}
\newcommand\AtPageUpperMyright[1]{\AtPageUpperLeft{%
 \put(\LenToUnit{0.5\paperwidth},\LenToUnit{-1cm}){%
     \parbox{0.5\textwidth}{\raggedleft\fontsize{9}{11}\selectfont #1}}%
 }}%
\newcommand{\conf}[1]{%
\AddToShipoutPictureBG*{%
\AtPageUpperMyright{#1}
}
}
\newcommand\copyrighttext{%
  \footnotesize \textcopyright 2024 IEEE. Personal use of this material is permitted.  Permission from IEEE must be obtained for all other uses, in any current or future media, including reprinting/republishing this material for advertising or promotional purposes, creating new collective works, for resale or redistribution to servers or lists, or reuse of any copyrighted component of this work in other works.}
\newcommand\copyrightnotice{%
\begin{tikzpicture}[remember picture,overlay]
\node[anchor=south,yshift=10pt] at (current page.south) {\fbox{\parbox{\dimexpr\textwidth-\fboxsep-\fboxrule\relax}{\copyrighttext}}};
\end{tikzpicture}%
}
\IEEEoverridecommandlockouts

\begin{document}

\title{AI-based Dynamic Schedule Calculation in Time Sensitive Networks using GCN-TD3\\

\thanks{This work has been funded by the Federal Ministry of Digital and Transport Germany under grant number 19OI22015F. The authors are responsible for the content of this paper.}
}

\author{\IEEEauthorblockN{Syed Tasnimul Islam\IEEEauthorrefmark{1}}
\IEEEauthorblockA{
\textit{TU Chemnitz}\\
Chemnitz, Germany \\
syed-tasnimul.islam@etit.tu-chemnitz.de}
\and
\IEEEauthorblockN{Anas Bin Muslim}
\IEEEauthorblockA{\textit{University of Applied Sciences Osnabrück}\\
Osnabrück, Germany\\
a.bin-muslim@hs-osnabrueck.de}
}

\maketitle
\conf{Accepted article IFIP/IEEE Networking 2024 (Tensor)
}

\copyrightnotice
\begin{abstract}
Offline scheduling in Time Sensitive Networking
(TSN) utilizing the Time Aware Shaper (TAS) facilitates optimal
deterministic latency and jitter-bounds calculation for Time-
Triggered (TT) flows. However, the dynamic nature of traffic
in industrial settings necessitates a strategy for adaptively 
scheduling flows without interrupting existing schedules. Our research identifies critical gaps in current dynamic scheduling methods for TSN and introduces the novel GCN-TD3 approach. This novel approach utilizes a Graph Convolutional Network (GCN) for representing the various relations
within different components of TSN and employs the Twin
Delayed Deep Deterministic Policy Gradient (TD3) algorithm
to dynamically schedule any incoming flow.  Additionally, an
Integer Linear Programming (ILP) based offline scheduler is
used both to initiate the simulation and serve as a fallback
mechanism. This mechanism is triggered to recalculate the entire
schedule when the predefined threshold of Gate Control List(GCL) length exceeds.
Comparative analyses demonstrate that GCN-TD3 outperforms
existing methods like Deep Double Q-Network (DDQN) and Deep
Deterministic Policy Gradient (DDPG), exhibiting convergence
within 4000 epochs with a 90\% dynamic TT flow admission rate
while maintaining deadlines and reducing jitter to as low as 2us.
Finally, two modules were developed for the OMNeT++ simulator,
facilitating dynamic simulation to evaluate the methodology.

\end{abstract}

\begin{IEEEkeywords}
Time Sensitive Network, TD3, Graph Convolutional Network.
\end{IEEEkeywords}

\IEEEpeerreviewmaketitle

\section{Introduction and Related Work}
Time-Sensitive Networking (TSN) is a Layer 2 technology, a collection of standards orchestrated by the IEEE 802.1 working group’s TSN task group. Originating from the enhancements to Audio Video Bridging (AVB) technologies, TSN standards are designed to enable precise time-synchronization, ensure bounded low latency, guarantee high reliability, and facilitate effective resource management. Among the versatile toolkits of TSN, central to this paper is the exploration of  IEEE 802.1Qbv, which plays a pivotal role in TSN. This standard employs GCL at the egress ports of switches, a crucial feature that enables the synchronized execution of distributed tasks across the network, thus meeting stringent real-time operational demands.

Current scheduling models in TSN predominantly rely on static methods, utilizing tools like Network Calculus, Satisfiability Modulo Theory (SMT) solvers and Integer Linear Programming (ILP) for ensuring bandwidth and latency guarantees \cite{survey_stber_2022}. Although these approaches are effective in small-scale scenarios, they often struggle to deliver timely solutions in dynamic environments,  within a reasonable time in a more dynamic context, such as manufacturing pipelines when the network and flow set size grows \cite{reinforcement_min_2023}. Literature review in \cite[Tab. 2]{survey_stber_2022} indicates the research gap of schedule calculation with a fixed routing in the domain of runtime reconfiguration of flows. Further literature review, summarized in Tab. \ref{table1}  reveals gaps in dynamic scheduling solutions, with studies either leaning towards non-scalable exact methods \cite{incremental_nayak_2018} or heuristic approaches  \cite{runtime_raagaard_2017}, \cite{runtime_li_2022}, \cite{dynamic_syed_2021}, \cite{adaptive_wang_2019}. Despite their attempt to address the computational challenge of NP-hard scheduling problems, these methods lack the balance between efficiency and adaptability. 

Recent advancements in Reinforcement Learning (RL) have introduced scalable and adaptable solutions for industrial automation and dynamic task rescheduling in TSN, employing algorithms like DDQN \cite{joint_yang_2022}, \cite{drls_zhong_2021}, Policy Gradient\cite{ttdeep_jia_2021}, and Asynchronous Advantage Actor-Critic methods\cite{deep_wang_2022}. However, the efficacy of these solutions is contingent upon the strategic selection of algorithms, failing to do so several limitations persist across these studies. Such as the inefficient segmentation of the time domain into fixed slots, i.e. \cite{ttdeep_jia_2021} sized relative to the Maximum Transmission Unit (MTU). Nevertheless, they disregard both the varied nature of time-critical scheduled flow size in industrial use cases\cite[Tab. 12]{ tsn_flexible_manufacturing_2018} and tick granularity (10ns) guideline from IEEE 802.3 working group hence waste network resources. Moreover, the reliance of DRL methods on Convolutional Neural Networks or Multi-Layer Perceptrons introduces challenges with data in non-Euclidean space, necessitating retraining for each new flow or device addition.

The interaction between an ideal RL agent and its environment, crucial for dynamic decision-making, is often hindered by the absence of appropriate simulation tools and cannot solely rely on synthetic data. Although various studies have utilized tools like Omnet++ \cite{drls_zhong_2021}, \cite{reconfiguration_nasrallah_2019} and Networkx \cite{runtime_li_2022}, \cite{online_yu_2020}, \cite{incremental_nayak_2018}, \cite{joint_yang_2022}, the lack of dynamic simulation capabilities or clear methodologies for integrating such features highlights a significant gap in current research approaches.

\begin{table*}[htbp]
\begin{threeparttable}
\centering
\caption{Overview of features of algorithms focusing on dynamic schedule calculation}
\label{table1}
\begin{tabular}{|l|l|l|l|l|l|l|l|l|l|l|l|l|l|l|*{14}{c|}}\hline
\backslashbox{Features}{Publications} & \cite{adaptive_wang_2019} & \cite{decentralized_polachan_2021} & \cite{deep_wang_2022} & \cite{dynamic_syed_2021} & \cite{incremental_nayak_2018} & \cite{joint_yang_2022} & \cite{leveraging_grtner_2021} & \cite{online_yu_2020} & \cite{reconfiguration_nasrallah_2019} & \cite{runtime_raagaard_2017} & \cite{runtime_li_2022} & \cite{ttdeep_jia_2021} & \cite{drls_zhong_2021} & \cite{reinforcement_min_2023} \\ \hline
Offline Initial Schedule & \fullcirc & - & - & \fullcirc & \fullcirc & - & - & \fullcirc & - & \fullcirc & \fullcirc & - & - & - \\ \hline
Mixed Traffic & - & \fullcirc & - & - & \fullcirc & \fullcirc & - & - & \fullcirc & \halfcirc & \fullcirc & - & - & - \\ \hline
GCL synthesis & - & - & \fullcirc & - & - & - & - & \fullcirc & - & \fullcirc & - & \fullcirc & \fullcirc & - \\ \hline
Dynamic Simulation & - & \fullcirc & - & - & \halfcirc & \halfcirc & - & \halfcirc & \halfcirc & - & \halfcirc & - & \halfcirc & - \\ \hline
Flow to Queue Mapping & - & \fullcirc & \fullcirc & - & - & \fullcirc & - & - & \fullcirc & \halfcirc & - & \fullcirc & \fullcirc & - \\ \hline
DynamicQueueAssignment & - & - & - & - & - & - & - & - & - & \fullcirc & - & - & - & - \\ \hline
Reschedule Trigger & \fullcirc & - & - & - & - & - & \fullcirc & \fullcirc & - & \fullcirc & - & - & - & - \\ \hline
Reinforcement Learning & - & - & \fullcirc & - & - & \fullcirc & - & - & - & - & - & \fullcirc & \fullcirc & \fullcirc \\ \hline
GNN & - & - & - & - & - & \fullcirc & - & - & - & - & - & \fullcirc & - & - \\ \hline
\end{tabular}
\begin{tablenotes}
\item \fullcirc: Feature Provided.  \halfcirc: Feature partially provided.   (-): Feature not provided.
\end{tablenotes}
\end{threeparttable}
\end{table*}
We now turn our attention to GCL length, queue usage, and flow-to-queue mapping in commercial TSN switches, notably those like Innoroute Trustnode \cite{innoroute}, limited by a GCL length threshold ($\omega$) of 128 entries per egress port. Craciunas et al. \cite{craciunas_scheduling_2016} thoroughly discussed the challenges of egress interleaving and flow isolation as per the 802.1Qbv standards, emphasizing the need for flow to queue assignment and queue management mechanism. The lack of verification against simulated switch models further exempts researchers from addressing these critical considerations. Finally, reconfiguration algorithms aim to integrate flows with minimal, deterministic response time but fall short of the efficiency of offline exact methods. This efficiency gap underscores an unavoidable reliance on the \textit{hyperperiod} (HP)—the maximum length of the time domain available for accommodating all flows—prompting the necessity for offline recalculations and system restarts to ensure optimal flow handling. Yet, few studies \cite{adaptive_wang_2019}, \cite{leveraging_grtner_2021}, \cite{online_yu_2020}, \cite{runtime_raagaard_2017} address this need for a fallback mechanism by halting runtime reconfiguration algorithms to recalculate and reintegrate all flows, highlighting a significant gap in dynamic TSN scheduling research.
Motivated by the identified research gaps, this paper introduces several key contributions:
\begin{itemize}
\item Integration of GCN \cite{Chen_Wang_Wang_Kuo_2020} with RL agent to enhance management of topology updates, incoming flows, and system state changes.
\item Employment of TD3, an off-policy actor-critic algorithm, to address time domain slot segmentation inefficiencies, offering improved sample efficiency and reduced bias compared to predecessors like DDQN and DDPG.
\item Enhancement of the INET framework within OMNET++ for dynamic simulations, providing a more realistic testing environment for TSN reconfigurations.
\item Development of a fallback mechanism that recalculates schedules offline using an ILP-based scheduler, considering GCL length constraints of switches.
\item Dynamic assignment of queues by algorithms based on real-time network conditions to boost flow handling and network efficiency.
\end{itemize}

\section{System Model}
In this study, we define Time Sensitive Networking (TSN) using standard terminology, incorporating \textit{End Systems} (ES) equipped with network interface cards (NIC) and TSN switches (or bridges, terms used interchangeably), interconnected by physical links and fully synchronized using the IEEE 802.1 AS protocol. Following IEEE 802.1Qcc, ESs are categorized as \textit{talkers} (TK) and \textit{listeners} (LR), with the assumption of deterministic traffic dispatch, potentially via tools like Intel’s DPDK. Our model assumes a store-and-forward mechanism in switches, simplifying the focus solely on transmission delays by omitting processing and propagation delays and inherently eliminating queuing delay due to our model's no-wait feature.

\begin{figure}[htbp]
\centering
\includegraphics[width=\linewidth]{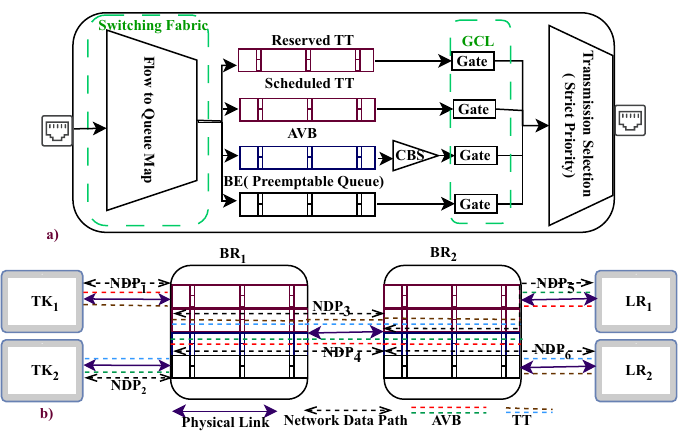}
\caption{(a) The proposed node model. (b) Pictorial representation of data path and network data path of 4 flows. }
\label{fig:SystemModel}
\end{figure}

The network topology is depicted as a directed graph, with vertices representing ESs, switch queues and edges symbolizing full-duplex links. Detailed topology modelling is discussed in Section \ref{sec:GCN}. We adapt the configuration 3 from \cite{craciunas_scheduling_2016} with a novel twist: we limit 4 total queues, three for static scheduling scenario and one reserved for dynamically arriving critical flows. Our strategy designates one queue for critical traffic, another for Audio Video Bridging (AVB) traffic, specifically audio, and a third for Best Effort (BE) traffic, as depicted in Fig. \ref{fig:SystemModel}(a). The intent to reserve an extra queue for dynamic scheduling scenario is to avoid egress interleaving problem \cite{craciunas_scheduling_2016}. Additionally to clarify further the leftover  unallocated timeslot for the  3 Scheduled Queues $Q_s\in (Q_{Reserved}\cap Q_{Scheduled}\cap Q_{AVB})$ is  allocated for the BE Queue ($Q_b$). Staying consistent with industrial use-cases scenario \cite{tsn_flexible_manufacturing_2018}, we assume all flow sizes are below the Maximum Transmission Unit (MTU), including AVB (audio traffic), as defined by IEEE working groups\cite{IEEE}. Flow routing calculations fall outside this paper's scope. In our model, each flow \(f_i \in F\) is defined by the tuple ${(D_i, L_i, P_i)}$, representing the deadline, duration (akin to size) and  periodicity, respectively. For a given flow \(i\) associated with bridge \(j\), the allocated queue is denoted as \(BR_j.Q_i\).    
Adopting the term \textit{datapath} from Pop. et al.\cite{pop_design_optimisation_2016}, denoted as \(dp_j\), which represents an ordered sequence of links from a talker \(TK_j \in TK\) to a listener \(LR_k \in LR\), we extend this concept to define the \textit{Network Datapath} (NDP). An NDP represents a subset of a datapath that requires a unique, non-overlapping schedule for each flow. Unlike traditional graph formulations that consider bridges merely as nodes, our approach uniquely identifies each bridge's distinct queues as separate nodes. For example, in Fig. \ref{fig:SystemModel}(b) a scenario comprising two (TT) and two Audio (AVB) flows the formulation of $NDP$ for one TT and one AVB flows is as follows: 
\begin{align*}
dp_{\text{Brown}}(TT) = [[TK_1, BR_1], [BR_1.Q_3, BR_2], [BR_2.Q_3, LR_2]] \\
                        = [NDP_1, NDP_3, NDP_6].\\
dp_{\text{Green}}(AVB) = [[TK_2, BR_1], [BR_1.Q_2, BR_2], [BR_2.Q_2, LR_1]]\\
= [NDP_2, NDP_4, NDP_5]
 \end{align*}

\section{Problem Formulation}\label{sec:Problem Formulation}
In this study, we adopt the novel concept of applying the No-wait Job Shop Scheduling (NW-JSSP) methodology to TSN scheduling adapting the name as \textit{No-wait Flow Shop Scheduling problem} (NW-FSSP), drawing inspiration from \cite{durr_nayak_2016}. 
We describe the dynamic NW-FSSP with random flow arrival using symbols defined as follows: There are $n$ successively arriving flows $F = \left\{ f_{1},f_{2},...,f_{n} \right\}$ which should be processed on $k$ NDPs. Each flow $F_i$ consists of a predetermined sequence of $h_i$ operations. $O_{if}$ is the $f^{th}$ operation of flow $F_i$, which can be processed on assigned NDPs. The processing time of $O_{if}$ on  $NDP_k$ is denoted by $T_{ik}$.
As, the minimal expectation from TAS is to have all the scheduled flows reach their destination within the deadline requirement, the objective here is to obtain flow offset $\phi_{i}$ of the dynamically arriving flow such that end-to-end latency of all scheduled flows are minimized.
\begin{equation}\label{ObjectiveFunction1}
 z = \min( \sum_{i=0}^{F}( \max(0, E_{ij} - D_{i}) \cdot w_{i}) )
\end{equation}
Here, $E_{ij}$ denotes the latency observed by flow $i$ in datapath $j$ and weight of the flow is equivalent to the Priority Code Point (PCP) value of the flow. The constraints that are considered both for ILP formulation and RL's action and reward design are as follows: 
\begin{enumerate}\label{constraints}
    \item A flow operation in a subsequent NDP will only commence after the completion of the previous flow operation if the flow traverses multiple NDPs.
    \item The end-to-end delay of a flow must not exceed the specified deadline requirement.
    \item The offset of a scheduled flow must be non-negative, and the entire transmission window must be accommodated within the flow's periodicity.
    \item Flows within the same NDP must not overlap temporally.
 \end{enumerate}
\section{GCN-TD3 FrameWork Overview}
\begin{figure}[htbp]
\centering
\includegraphics[width=\linewidth]{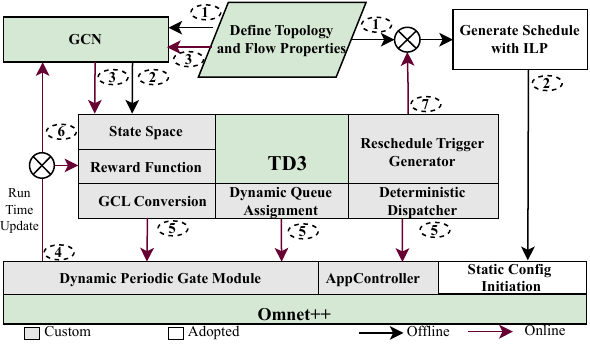}
\caption{GCN-TD3 Framework overview.}
\label{fig:FrameWork}
\end{figure}

In Fig.~\ref{fig:FrameWork}, we present our novel GCN-TD3 framework, designed as a  runtime scheduling mechanism for dynamic flow handling in TSN. Our experimentation utilizes the INET Framework of discrete event simulator Omnet++ and two custom modules to facilitate dynamic simulation. The simulation initiates (Step 1) with the importation of a predefined network topology and flow properties from a CSV file. These are initially processed by the GCN and ILP generator modules. Subsequently, in Step 2, the ILP module constructs the initial Gate Control List (GCL), which is then forwarded to Omnet++ to kickstart the simulation, while the GCN module establishes the state space necessary for the TD3 algorithm, as described in Section \ref{sec:GCN}. In Step 3, to simulate the arrival of dynamic flows, flow properties are randomly fetched from the CSV file and processed by the GCN to generate state updates for TD3.
 In Step 4, guided by Algorithm \ref{alg:TD3} and incorporating runtime system data, TD3 decides on the integration of new flows. By Step 5, the custom \textit{AppController},  module begins its role as the first dynamic module, inheriting from the cSimpleModule class of Omnet++. It periodically retrieves data from the \textit{Deterministic Dispatcher} module, such as node name, application index, and initial offset and adjusts application parameters within the INET hierarchy using the getAncestorPar() function. Concurrently, the \textit{DynamicPeriodicGate}  module, extending INET's \textit{PeriodicGate} module, reads  CSV file generated by \textit{GCL Conversion}  module to update the GCL  for specified transmission gates, seamlessly integrating dynamic flow changes into the simulation. TD3 also calculates a reward function based on real-time simulation data in Step 6. Should TD3 reject a flow, Step 7 is initiated to evaluate the need for a complete rescheduling through the static ILP scheduler, triggering a fallback mechanism for an offline recalibration with ILP and a process restart if deemed necessary. 
\subsection{Generate Static Schedule with ILP}

The detailed ILP formulation, tailored to the constraints outlined in our problem definition, is thoroughly documented in works of Pop et al.\cite{pop_design_optimisation_2016} and Craciunas et al. \cite{craciunas_scheduling_2016}.
We recommend readers consult these sources for a comprehensive  understanding. Given our paper's emphasis on dynamic flow scheduling, we have adapted the mathematical models from these references, applying them both for initial static scheduling and as a fallback recalculation strategy in scenarios where dynamic scheduling encounters challenges.

\subsection{Graph representation learning/embedding with GCN}\label{sec:GCN}
\begin{figure}[htbp]
\includegraphics[width=\linewidth]{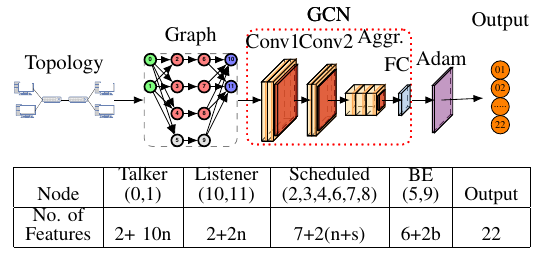}
\caption{Process of fixed size representation of variable size topology and flow features by GCN.}
\label{fig:GCN}
\end{figure}

In this study, we overcome the challenge posed by variable-sized node features in a TSN topology by employing a GCN model. The topology is modelled as a graph \( G(\mathbf{A}) = (V, E) \) where nodes \( V \) represent entities like talkers, listeners, scheduled queues, and BE queues in switches. Each node type possesses distinct and variable features. The graph structure is defined by an adjacency matrix \( \mathbf{A} \in \mathbb{R}^{N \times N} \), maps the interconnections among $N$ nodes with directed edges \( e_{ij} \in E \) denoting link from node \( i \) to node \( j \). Drawing inspiration from \cite{KipfW17}, our GCN model is specifically designed to manage these variable-sized features by initially passing them through specialized input layers designed for each node type. As shown in   Fig.~\ref{fig:GCN}, the architecture comprises two graph convolutional layers with 64 (Conv1)  and 32 ( Conv2) neurons respectively, using ReLU activation functions. 
These layers aggregate feature information from neighbouring nodes, encapsulating each node's hidden representation. Following this aggregation, a non-linear transformation refines the outputs and a Global Average Pooling layer (FC) reduces these node embeddings into a singular, unified graph representation. The training of our model is conducted using the ADAM optimizer, and a Mean Squared Error (MSE) loss function to fine-tune the model focusing on minimizing the error between the predicted and actual graph representations. Table \ref{tablegnn} represents features considered for each node type. Additionally, a summarized table presented at the bottom of Fig.~\ref{fig:GCN} offers a generalized overview of the feature size variations across different node types. This environment manages \(n\) flows, containing both Scheduled Queues (TT and AVB) and Best Effort (BE) queues. Each of these queues is characterized by \(s\) and \(b\) pairs of gate events respectively, which denote the specified gate opening and closing times within a GCL at each egress port. The decision to categorize TT and AVB queues together under scheduled queues, despite their different properties, stems from our dynamic scheduler’s strategy to integrate any new TT flow into the $HP$ seamlessly, ensuring there is no disruption to the existing schedule of both TT and AVB flows. In contrast, BE flows remain susceptible to preemption.

\subsection{TD3 algorithm for dynamic scheduling}

In this section, we reframe the problem outlined in Section \ref{sec:Problem Formulation} as a Markov Decision Process (MDP). The MDP is defined by the quartet ( $\mathcal{S}$,$\mathcal{A}$,$\mathcal{S’}$,$\mathcal{R}$), representing state, action, subsequent state, and reward, respectively. 
\subsubsection{State Space}:
 The state space $\mathcal{S}$ is a 22-dimensional vector, where each state \(s_i \in \mathcal{S}\) encapsulates the embedded graph representation of the network topology along with features as shown in Fig.\ref{fig:GCN}. Noteworthy features extracted from Omnet++ runtime are $E_i$, number of contained ($F_c$) and dropped flows ($F_d$) by individual queues highlighted in blue in Table \ref{tablegnn}. 
\subsubsection{Action Space}
The action \(a_t \in \mathcal{A}\) at time $t$ is defined as: 
$a_t = [a_1(t),a_2(t), a_3(t),..,a_{3+(mn-1)}(t), a_{3+mn}(t)]$; where $m$ = no. of switches and $n=HP/max(P_i)$ phase of a flow or sub-cycle and Inter Frame Gap ($IFG$) is 12 B.
\begin{itemize}\label{item:action}
    \item \( a_1 \in \{0, 1\} \): Reject or accept the incoming  flow.
    \item  \( a_2 \in \{0, 1\} \): Forward to scheduled or reserved Queue. 
   \item  \( a_3 \in [0, (HP/n)-(m+1)\times L_i- m\times IFG] \)\:: Deterministic dispatch time from talker.
   \item\( a_{3+(mn-1)} \in [a_3 + (n-1) \times P_i+ L_i+IFG, n \times P_i - (m \times L_i)] \): Gate opening time of the first bridge on the path. 
   \item \( a_{3+mn} \in [a_{3+(mn-1)} + L_i+IFG, n \times P_i -L_i] \):Gate opening time of the second bridge.
\end{itemize}
\begin{table}[htbp]
\begin{threeparttable}
\centering
\caption{Overview of node features for GCN}
\label{tablegnn}
\small\addtolength{\tabcolsep}{-5pt}
\begin{tabular}{|c|c|c|c|c|c|}
\hline
\multicolumn{2}{|l|}{\backslashbox{Features}{Node Type}} & Talker & Listener & \begin{tabular}[c]{@{}l@{}}Scheduled \\ Queue\end{tabular} & \begin{tabular}[c]{@{}l@{}}BE \\ Queue\end{tabular} \\
\hline
\multicolumn{2}{|l|}{ID} &\fullcirc&\fullcirc  & \fullcirc & \fullcirc\\ \hline
\multirow{10}{*}{\rotatebox{90}{No. of  Flows}} & PCP &\fullcirc &- & -& -\\ \cline{2-6}
& Flow ID & \fullcirc&\fullcirc &- & -\\ \cline{2-6}
& Size & \fullcirc&- & -&- \\ \cline{2-6}
& Period&\fullcirc & - & -&- \\ \cline{2-6}
& \begin{tabular}[c]{@{}l@{}}Latency \\ Bound\end{tabular} &\fullcirc & -&- & -\\ \cline{2-6}
& Destination & \fullcirc& -&- &- \\ \cline{2-6}
& Source & -& \fullcirc &- &- \\ \cline{2-6}
& Start Time & \fullcirc&- &\fullcirc &- \\ \cline{2-6}
& End Time &\fullcirc &- & \fullcirc& -\\ \cline{2-6}
& \begin{tabular}[c]{@{}l@{}}Bridge  \\ Delay\end{tabular} & \fullcirc&- &- &- \\ \cline{2-6}
&  \begin{tabular}[c]{@{}l@{}}\textcolor{blue}{Observed} \\ \textcolor{blue}{Latency}\end{tabular} & \fullcirc& -& -&- \\ \hline

\multicolumn{2}{|l|}{\textcolor{blue}{Flows Contained}} &- & - & \fullcirc &\fullcirc\\ \hline
\multicolumn{2}{|l|}{\textcolor{blue}{Flows Dropped}} &- & - & \fullcirc &\fullcirc\\ \hline
\multicolumn{2}{|l|}{Switch ID} &- & - & \fullcirc &\fullcirc\\ \hline
\multicolumn{2}{|l|}{Queue ID} &-& - & \fullcirc & \fullcirc\\ \hline
\multicolumn{2}{|l|}{Preemptivity} &- &  -& \fullcirc & \fullcirc\\ \hline
\multirow{2}{*}{\begin{tabular}[c]{@{}l@{}}Gate Event \\ Count\end{tabular}} & Open Time &- & -&\fullcirc &\fullcirc \\ \cline{2-6}
& Close Time  &- & -&\fullcirc &\fullcirc \\ \hline
\end{tabular}
\begin{tablenotes}
\item \fullcirc: Feature considered.    (-): Feature not considered.
\end{tablenotes}
\end{threeparttable}
\end{table}
\subsubsection{Reward Function}
The immediate reward formula:
\begin{equation}\label{eq:r}
r(s_t,a_t) = r(O).r(S).r(B).a_1 - \frac{\beta}{\omega} - \frac{F_d}{F_d+F_c} - \frac{J_\sigma}{J_\mu} 
\end{equation}
\begin{equation}
\label{eq:r(O)}
r(O) = 
\begin{cases} 
0, & \textbf{if } \exists f_i \in F \text{ such that } E_{i} - \phi_i > D_i \\
1, & \textbf{otherwise}
\end{cases}
\end{equation}
\begin{equation}
\label{eq:r(S)}
r(S) = 
\begin{cases} 
0, & \begin{array}{@{}l@{\quad}l}
\textbf{if } \exists f_r \in F, \exists q_s \in Q_s, \exists (o_{se}, c_{se}) \in E_{q_s} \\
\text{such that } (\phi_{rj} < c_{se}) \land ( \phi_{rj} + L_r > o_{se})
\end{array} \\
1, & \textbf{otherwise} 
\end{cases}
\end{equation}
\begin{equation}
\label{eq:r(B)}
r(B) = 
\begin{cases} 
1, & \begin{array}{@{}l@{\quad}l}
\textbf{if } \exists f_r \in F, \exists q_b \in Q_b, \exists (o_{be}, c_{be}) \in E_{q_b} \\
\text{such that } \\ (\phi_{rj} < c_{be} - L_g) \land (\phi_{rj} + L_r > o_{be} + L_g)
\end{array} \\
0, & \textbf{otherwise}  
\end{cases}
\end{equation}
The leftmost part of the reward function in \eqref{eq:r} ensures compliance with several critical constraints: it verifies upon acceptance of the new flow \(r\) whether all flows $f_i \in F$ reach within deadline \eqref{eq:r(O)}, prevents \(r\) from being scheduled in switch \(j\) during the time slots allocated to existing flows \eqref{eq:r(S)}, and ensures adherence to the preemption guard duration \(L_g\) requirements of IEEE 802.1 Qbu \eqref{eq:r(B)} when occupying time slots left for BE flows. The variables \(o_{se}, o_{be}\) and \(c_{se}, c_{be}\) denote the opening and closing times for gate events of the scheduled queues \(E_{qs}\) and BE queue \(E_{qb}\) respectively.
Penalties are applied to the agent if there's an increase in the GCL length counter \(\beta\), the number of dropped flows \(F_d\) across queues and  the coefficient of variation of jitter $\frac{J_\sigma}{J_\mu}$.

\subsubsection{Policy Optimization with TD3 algorithm}\label{sec:training}
As outlined in Algorithm \ref{alg:TD3}, first the policy parameter $\phi$ and the Q-function parameters $\theta_1$ and $\theta_2$ are initialized. Additionally, discounted reward factor $\gamma$, the frequency of delayed policy updates $\eta$, the proportion factor of target network updates $\tau$ and $\beta$ are also initialized.

For each iteration $t$, the agent explores the environment by selecting actions based on its current policy, adding some noise (as stated in line~\ref{lst:targetaction}) for exploration purposes. During the gradient step, a batch of experiences is randomly sampled from the stored replay buffer $B$ and used to update the Q-function, which evaluates the quality of actions taken by the agent. As stated in line~\ref{lst:bellman}, the Bellman equation\cite{fujimoto2018addressing} is used to calculate a target value where the smaller of the two outputs of target critic networks is fed into the equation to avoid overestimating the Q-value. The policy is then updated in a direction that maximizes the Q-function, indicating better action choices. The parameters of the target networks are not updated consecutively like those of the actor and critic networks to avoid overestimation during the training process. These parameters are updated after timestep $\eta$ and according to hyperparameter $\tau$ as shown in  line~\ref{lst:target1} and line~\ref{lst:target2}. The algorithm continually refines its policy through iteration, aiming to pinpoint the optimal action parameters for the agent within its environment until $\beta = \omega$. Then the \textbf{Reschedule Trigger Generator} module halts the algorithm and falls back to the offline scheduler.

\begin{algorithm}[htbp]
\caption{TD3 algorithm}
\label{alg:TD3}
\begin{algorithmic}[1]
\STATE \textbf{Initialize}:$\phi$, $\theta_1$, $\theta_2$, $\mathcal{B}$, $\eta$, $\gamma$,$\tau$ and $\beta$
\STATE where $\eta, \gamma, \tau \in$ [0,1].
\STATE\textbf{Set}  $\theta'_1 \leftarrow \theta_1$, $\theta'_2 \leftarrow \theta_2$, $\phi' \leftarrow \phi$.
\REPEAT
    \STATE Observe $s$ and select $a \sim  \pi_{\phi}(s_t) + \epsilon$,  $\epsilon \sim \mathcal{N}(0, \sigma)$
    \STATE Execute $a$ and observe next state $s'$ and reward $r$
    \STATE Store transition $(s, a, r, s')$ in $\mathcal{B}$
        \FOR{$t = 1, \dots, \chi$}
            \STATE Sample mini-batch of $N$ transitions from $\mathcal{B}$
            \STATE Compute target actions:
            \STATE $\tilde{a} \leftarrow \pi_{\phi'}(s') + \text{clip}(\mathcal{N}(0, \tilde{\sigma}), -c, c)$\label{lst:targetaction}
            \STATE Compute targets:
            \STATE $y \leftarrow r + \gamma(1-d) \min_{i=1,2} Q_{\theta_i'}(s',\tilde{a})$\label{lst:bellman}
            \STATE Update critics:
            \STATE $\theta_i \leftarrow \text{argmin}_{\theta_i} N^{-1} \sum (y - Q_{\theta_i}(s, a))^2$
            \IF{$t \mod \eta= 0$}
                \STATE Update policy: $\nabla_{\phi} J(\phi)$
                \STATE $ = N^{-1} \sum \nabla_a Q_{\theta_1}(s, a)|_{a=\pi_{\phi}(s)} \nabla_{\phi} \pi_{\phi}(s)$
                \STATE Update target networks:
                \STATE $\theta_i' \leftarrow \tau \theta_i + (1 - \tau) \theta_i'$\label{lst:target1}
                 \STATE  $\phi' \leftarrow \tau \phi + (1 - \tau) \phi'$\label{lst:target2}

            \ENDIF
        \ENDFOR
\UNTIL{$\beta = \omega$}
\end{algorithmic} 
\end{algorithm}

\section{Experimental Evaluation and Result Analysis}
We trained the GCN-TD3 algorithm on a robust setup: a virtual machine powered by an Intel Xeon CPU, 32 GB of RAM, Ubuntu 22.04 LTS, and an Nvidia Quadro RTX GPU, running TensorFlow 2.0. 
\begin{figure}[htbp]
\centering
\includegraphics[width=\linewidth]{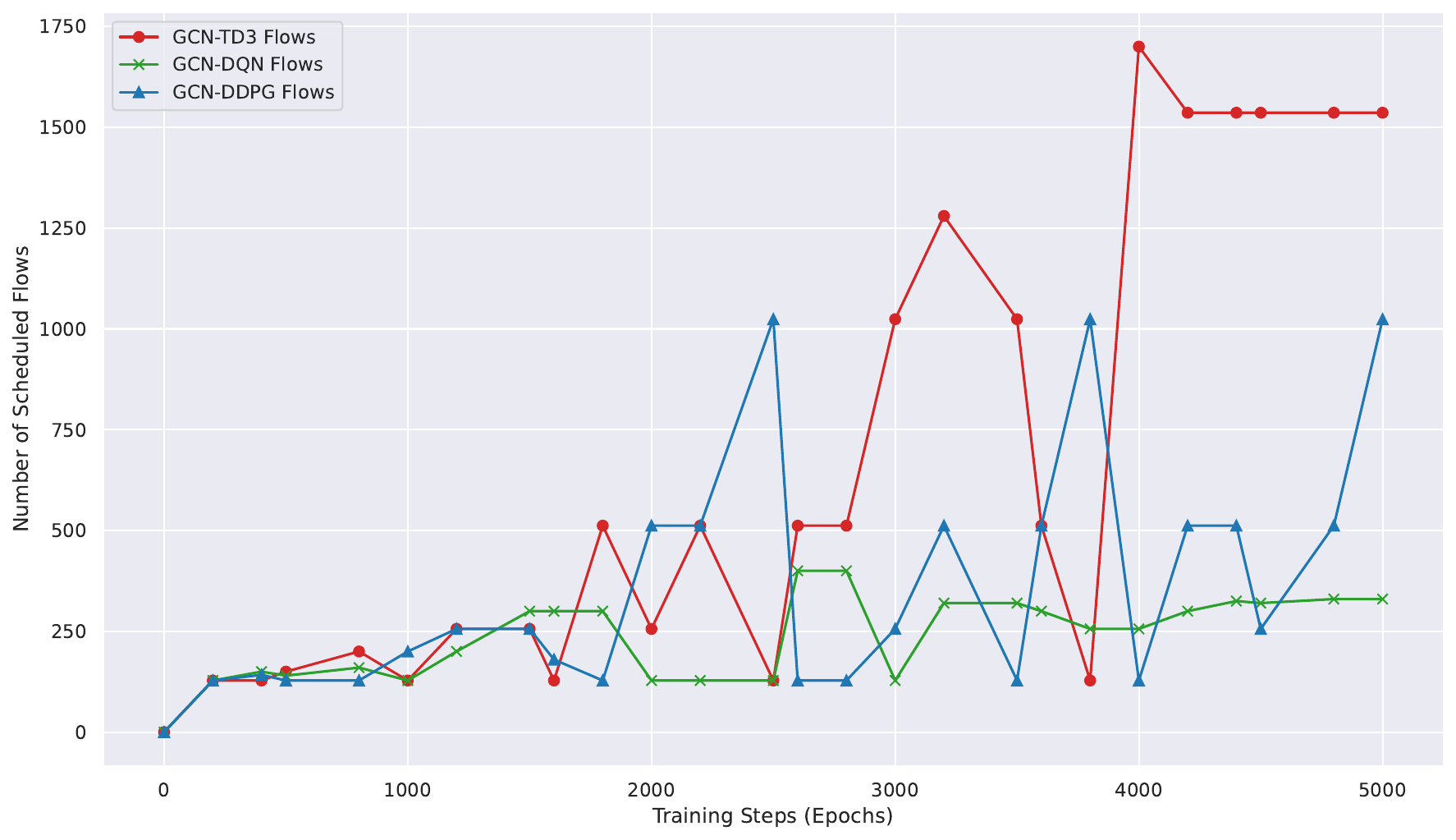}
\caption{No. of Scheduled Flows over 5000 training steps.}
\label{fig:result1}
\end{figure}

\begin{figure}[htbp]
\centering
\includegraphics[width=\linewidth]{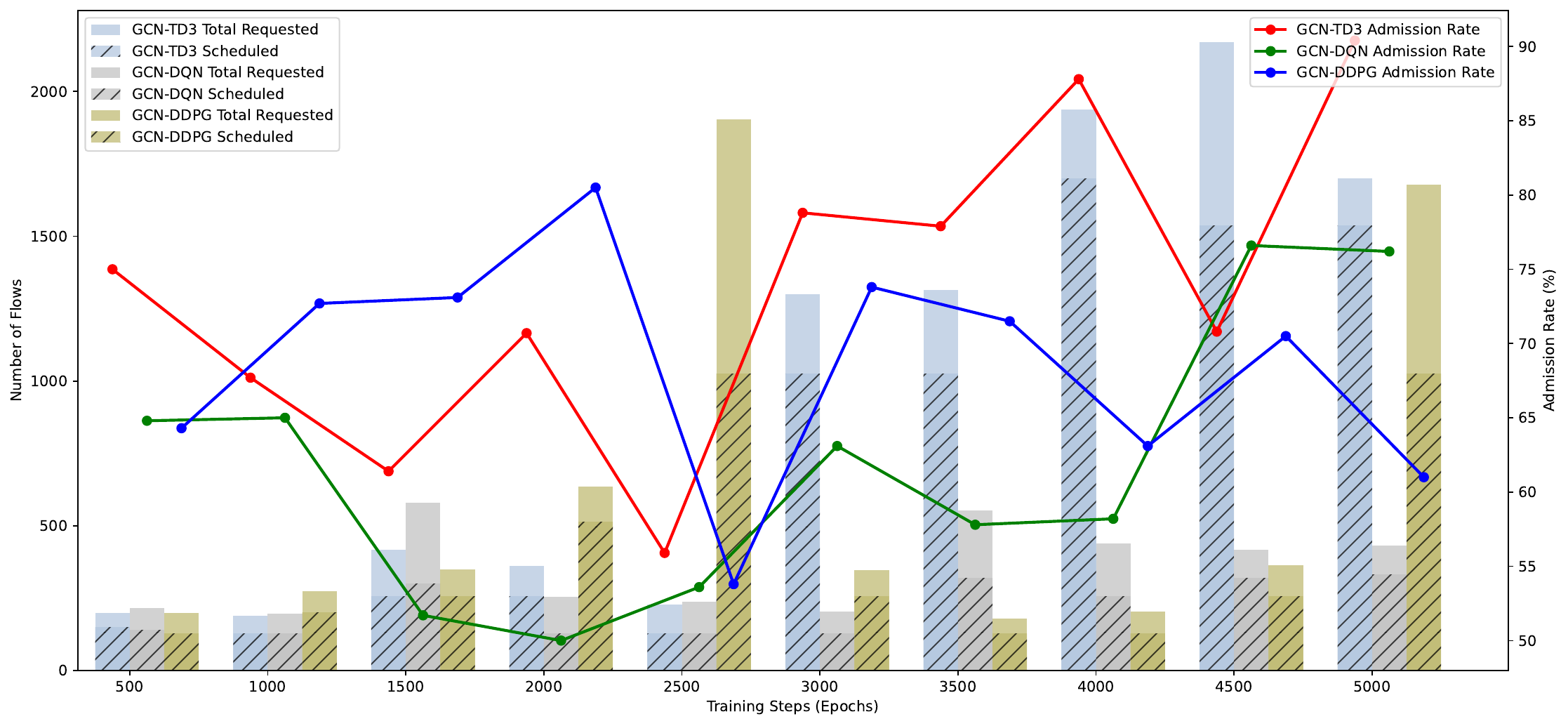}
\caption{Flow admission rate over 5000 training steps.}
\label{fig:result2}
\end{figure}
For solving integer linear programming (ILP), we turned to the academic version of the CPLEX solver. The training configuration for the TD3 agent involved statically defined source and destination nodes for each flow, with TT flows randomly selected based on the tuple $(Period, Size) = ((50B, 250us), (100B, 500us), (500B, 1ms))$, each having a 1 ms deadline. Two AVB flows of size {1000B, 1200B} and 4 ms period and 2 ms deadline, were scheduled statically alongside 2 TT flows using ILP. TD3's hyperparameters were aligned with those recommended in  \cite[Tab. 3 ]{fujimoto2018addressing}, with \(\omega\) set to 128. Over 5000 epochs, the algorithm was trained, with each epoch determined by \(\beta = \omega\). Performance metrics tracked per epoch included the total and requested number of flows, standard deviation of jitter, average end-to-end latency and flow admission rate.

Our study adopts a fully centralized configuration model as outlined by IEEE 802.1 Qcc, utilizing powerful servers in Centralized Network Configuration to run scheduling algorithms. These schedules are then communicated to TSN bridges via standard NETCONF protocol. The primary hardware limitation considered is the GCL length, which significantly influences our evaluation of scheduling algorithms to demonstrate the feasibility of the algorithms on commercial switches. To benchmark the effectiveness of GCN-TD3, it was compared against GCN-DDQN, which employs a time domain segmentation strategy typical in dynamic scheduling RL research \cite{reinforcement_min_2023}, \cite{ttdeep_jia_2021}, and DDPG, favoured by several studies for its policy gradient approach \cite{deep_wang_2022}. The performance comparison, depicted in Fig.\ref{fig:result1}, highlights GCN-TD3’s superior and stable flow accommodation capability within the constraints of GCL length. Although GCN-DDPG initially outperforms GCN-TD3, it fails to converge within the allotted time.  

\begin{figure}[htbp]
\centering
\includegraphics[width=\linewidth]{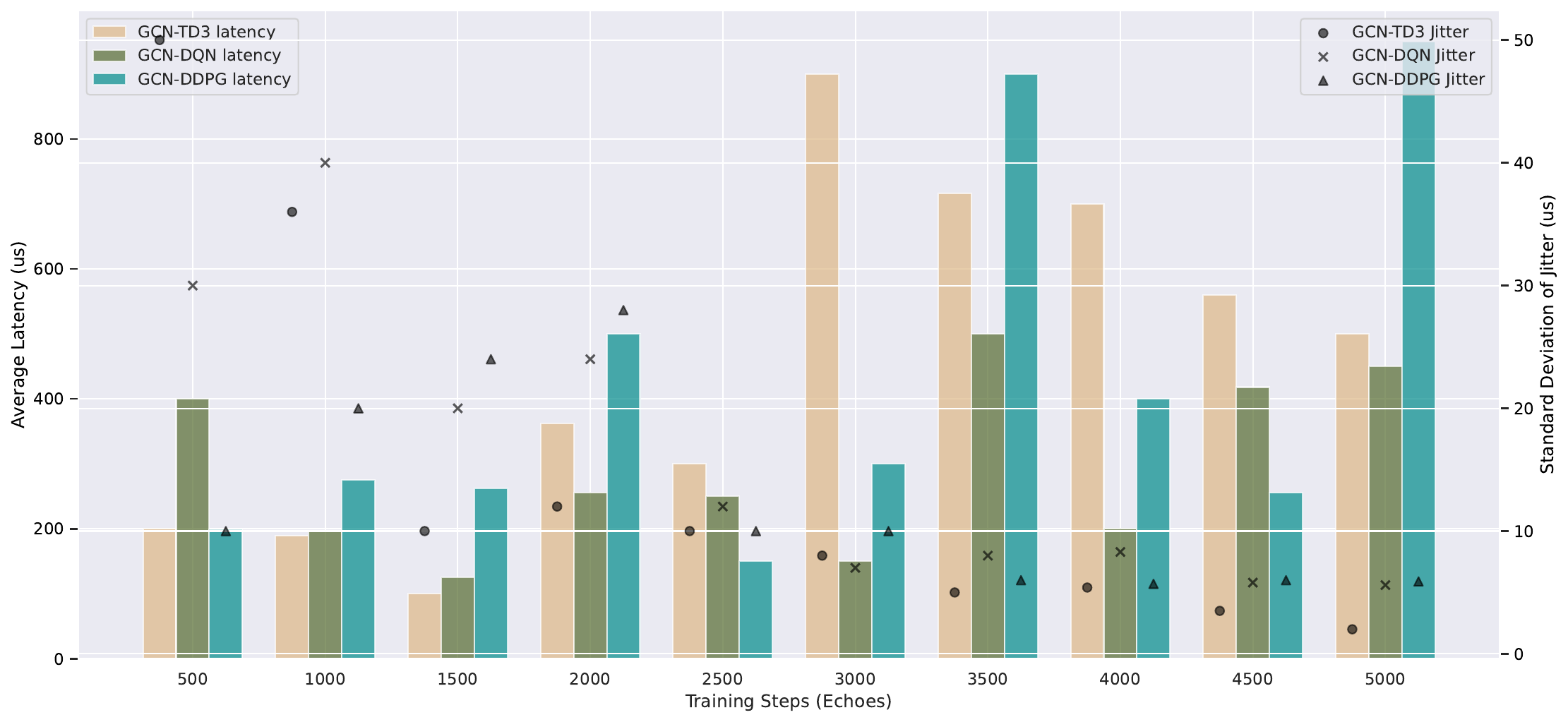}
\caption{Average Latency and Standard Deviation of Jitter over 5000 training steps.}
\label{fig:result3}
\end{figure}
It is important to consider metrics beyond just flow numbers for comprehensive evaluation, as shown in Fig.\ref{fig:result2}, where GCN-TD3 nearly achieves a 90\% admission rate. Even though GCN-DDQN had impressive admission rates, its total flow capacity was lower due to the inherent limitations of time domain segmentation. A key objective of our study is to minimize jitter while adhering to deadlines. As depicted in Fig.\ref{fig:result3}, although all algorithms initially struggled to balance jitter and latency, GCN-TD3 eventually achieved the lowest jitter (nearly 2us) while maintaining low latency and accommodating the most flows among the algorithms compared.
It is noteworthy that our initial intention was to evaluate the models using real network data in addition to doing simulation. However, as highlighted in \cite{ergenç2023synthesizing}, there is a lack of publicly available datasets for TSN that can be used for evaluating or benchmarking algorithms. Consequently, we plan to test the algorithm in future in our own TSN testbed once the setup is completed.
\section{Conclusion and Future Work}
In this paper, we introduce the GCN-TD3 framework, a cutting-edge AI-based solution designed for dynamic flow scheduling. This robust framework is carefully designed to align with the stringent requirements of deployable hardware and the intricate dynamics of network flows. Our extensive testing demonstrates that GCN-TD3 outperforms current state-of-the-art Reinforcement Learning methods, scheduling 30\% more TT flows and reducing jitter by 3\%.

However, the GCN-TD3 framework does have its limitations. It currently does not support dynamic path determination for flows and necessitates retraining with any change in the number of switches. Future research will aim to address these challenges by developing capabilities for dynamic routing and increasing reliability through the incorporation of redundant paths. Additionally, we plan to validate our findings within a physical testbed to facilitate the implementation of a digital twin system, further proving the effectiveness and applicability of our framework in real-world scenarios.

\bibliographystyle{IEEEtran}
\bibliography{refs}

\end{document}